# Multi-functional interface between integrated photonics and free space

Quentin A. A. Tanguy[1, §], Arnab Manna[2, §], Saswata Mukherjee[1, §], David Sharp[2, §], Elyas Bayati[1], Karl F. Böhringer[1,3], and Arka Majumdar[1,2]

[1]Department of Electrical and Computer Engineering, University of Washington, Seattle, WA, USA

[2]Department of Physics, University of Washington, Seattle, WA, USA

[3]Institute for Nano-engineered Systems, University of Washington, Seattle, WA, USA

[§]Equal contribution

* Corresponding author: arka@uw.edu

## Abstract

The combination of photonic integrated circuits and free-space meta-optics has the ability to unclasp technological knots that require advanced light manipulation due their conjoined ability to guide and shape electromagnetic waves. The need for large scale access and component interchangeability is essential for rapid prototyping of optical systems. Such capability represents a functional challenge in terms of fabrication and alignment of compound photonic platform. Here, we report a multi-functional interface that demonstrates the capabilities of a flexible and interchangeable combination of a photonic integrated circuit to a free-space coupling chip with different designs of low-loss meta-optics at a wavelength of 780 nm. We show that robustness and fidelity of the designed optical functions can be achieved without prior precise characterization of the free-space input nor stringent alignment between the photonic integrated chip and the meta-optics chip. A diffraction limited spot of approximately 3 µm for a hyperboloid metalens of numerical aperture 0.15 was achieved despite an input Gaussian elliptical deformation of up to 35% and misalignments of the components of up to 20 µm. A holographic display with a peak signal-to-noise ratio of more than 10 compared to its ground truth is also reported using this platform making this work the first interface to shape photonic integrated modes into free space using different diffractive optical functions.

# Introduction

With the rise of computational electromagnetics techniques and sophisticated nanofabrication technologies, we can now design and fabricate dielectric photonic structures, with unprecedented reduction in size, and weight, while simultaneously increasing their functionalities. These photonic structures are largely of two types: on-chip photonic integrated circuits (PIC), where the light is guided on the chip using waveguides and resonators, and sub-wavelength patterned diffractive optics, also known as meta-optics that shape the optical wavefront in free space[1]. The PIC and free-space optics serve two very different purposes: while free-space optics boast very large space-bandwidth product[2], as well as flexibility and compatibility with existing macroscopic optics[3,4], PICs can enable strong light-matter interaction due to tight spatial confinement of light and longer propagation distance via waveguiding. However, to create a functional compact optical system, we almost always need both free-space macroscopic optics and PICs. To interface PICs with free-space optics, we generally rely on grating couplers[5], which have limited functionality for spatial mode shaping. Thanks to the ability of meta-optics to arbitrarily shape the optical wavefront[6], we can manipulate the output light from the grating couplers to realize a multi-functional interface between PICs and free-space optics. Light coming out of a PIC will be shaped for desired functionalities using different meta-optics, and the wavefront-shaped light will be further routed via additional free-space optics. Such hybrid meta-optics PIC systems will be packaged together and can enable drastically miniaturized optical systems, with potential applications in beam steering[7–10], generation of structured light[11], optical trapping and manipulation of cold atom qubits[12–14]. To that end, researchers have already demonstrated focusing of light coupled from a PIC using meta-optics[7,10,15–18]. Additionally, reseachers have demonstrated holography by fabricating meta-optics on top of a PIC[15,17,19]. However, no multi-functional coupling between PIC and free space has yet been reported. Here, we report a chip-scale hybrid nanophotonic platform consisting of a two-dimensional array of identical apodized gratings on a PIC and an array of different meta-optics for shaping light coming out of each grating in a pre-defined fashion (Fig. 1a, b). The PIC and the meta-optics chips are separately fabricated and aligned. We tested the PIC/meta-optics platform at a wavelength of $\lambda=780$ nm, and demonstrated different optical functions, including, conventional lensing, extended depth of focusing, vortex beam generation and holographic projections. In total, we demonstrated simultaneous feeding of light from 14 apodized gratings in a PIC to 14 different meta-optics. Our work shows the ability of such a hybrid PIC/meta-optical platform to create multi-functional optical beams in parallel. Importantly, we showed that such a platform does not require stringent alignment (within up to 20 μm shift measured from the center of the apodized grating). We emphasize that the demonstrated functionalities in this paper are chosen as just some examples to highlight the capabilities of this platform. Depending on specific applications, different meta-optics can be designed.

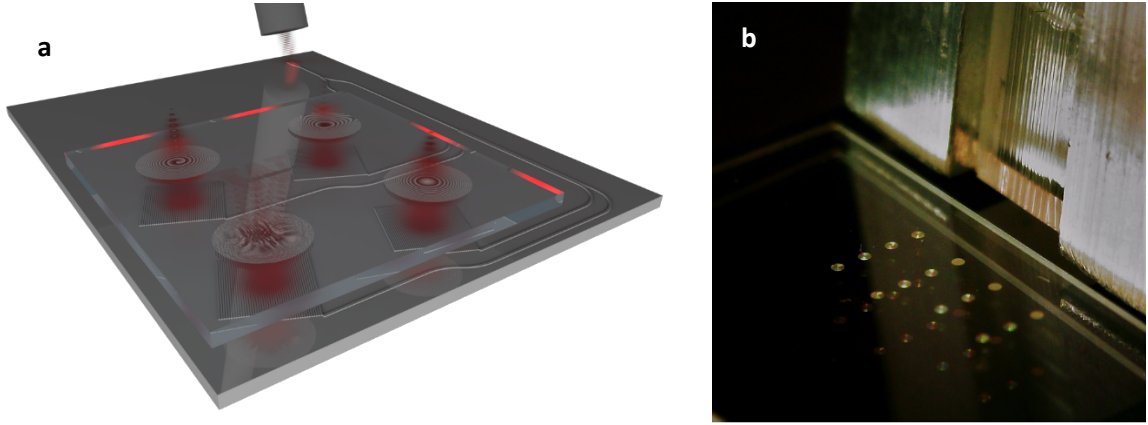

**Figure 1: Hybrid PIC/meta-optical system. a** Schematic of the system: an optical fiber feeds the laser light simultaneously to all the apodized gratings in a PIC via waveguides. The light comes out of the gratings and a separate chip containing an array of meta-optics shapes the grating-output in parallel. **b** Close-up image of the entire hybrid platform with the array of 14 meta-optics, placed on top of the PIC (covered by the meta-optics chip). The optical fiber array shown in the right is used to feed laser light to the PIC.

# Results

A schematic of the hybrid photonic platform is shown in Fig. 1a, showing different meta-optics (fabricated in a thin film of silicon nitride on the top of a quartz slab) aligned to an array of apodized gratings fabricated in the bottom PIC. We first design 300 μm-aperture apodized gratings following the approach by Kim *et al.*[20] to output a near-Gaussian beam to free space. These Gaussian beams are then shaped via meta-optics. Figure 1b shows the fabricated device in the optical characterization setup. It consists of an array of 16 compound devices (4×4 array) with identical apodized gratings (an optical microscope image shown in Fig. 2a) and different meta-optics sitting atop of the gratings. Because of fabrication imperfections, two apodized gratings were damaged and could not be probed. Out of the 14 functioning devices (Fig. 1b), 8 were retained for extensive characterization.

The photonic waveguides and gratings are made in $Si_3N_4$ to minimize the losses at the design wavelength of 780 nm. The 16 apodized gratings are connected through a single-entry port grating coupler, requiring a single optical fiber for the integrated photonic chip to function, splitting the input power near-equally between all of them. The wavefront generated in free space was purposefully not quantitatively characterized prior to the design of the meta-optics to demonstrate the robustness of the hybrid platform in terms of ability to reshape the wavefront despite the discrepancies between the design and the real experimental conditions. Optical and scanning electron microscopes pictures of an apodized grating are shown in Fig. 2b and 2c, respectively. The angle of the apodized grating relative to the waveguide $\theta_{inc}$ was chosen around 24° to minimize the divergence of the collimated Gaussian beam. *A posteriori* characterization of the Gaussian intensity on the grating plane was carried out and is shown in Supplement Fig. S1. A two-dimensional Gaussian fit on the measured intensity of the grating provides a waist of 98 μm along $\vec{x}$ and 137 μm along $\vec{y}$, which is respectively 35% and 9% smaller than the target widths. The center of the Gaussian beam is shifted off the center of the grating by approximately 23 μm on the x axis and -2 μm on the y axis.

The meta-optics were designed without prior characterization of the PIC. To carry out the design of the PIC and the meta-optics separately, we assumed that an accurate prior knowledge of the free-space field was not necessary to design the phase profiles. In particular, we made three assumptions based on previous studies[20], that allowed us to treat the incident field on the meta-optics as a centered plane wave with Gaussian intensity. First, the wavefront error of the Gaussian beam propagating to free space is sufficiently small to be considered a plane wave. Second, the out-coupling angles, defined from the $\vec{z}$ axis, orthogonal to the plane of the grating (see Fig. 2b) are also negligible[16]. And third, the free-space beam intensity is considered to have a regular (non-elliptical) Gaussian distribution. The details of the design and fabrication of the meta-optics are provided in Materials and Methods. We tested 14 different meta-optics and presented detailed results for eight of them: five meta-lenses (one with hyperboloid phase profile, two vortex beam generators with orbital angular momenta (*l*) of 3 and 15, and two extended depth of focus lenses with logarithmic-asphere and cubic phase profiles) and three holograms. The aperture of the meta-optics is kept at 300 μm to match that of the apodized gratings, although same aperture size is not crucial for the demonstrated platform.

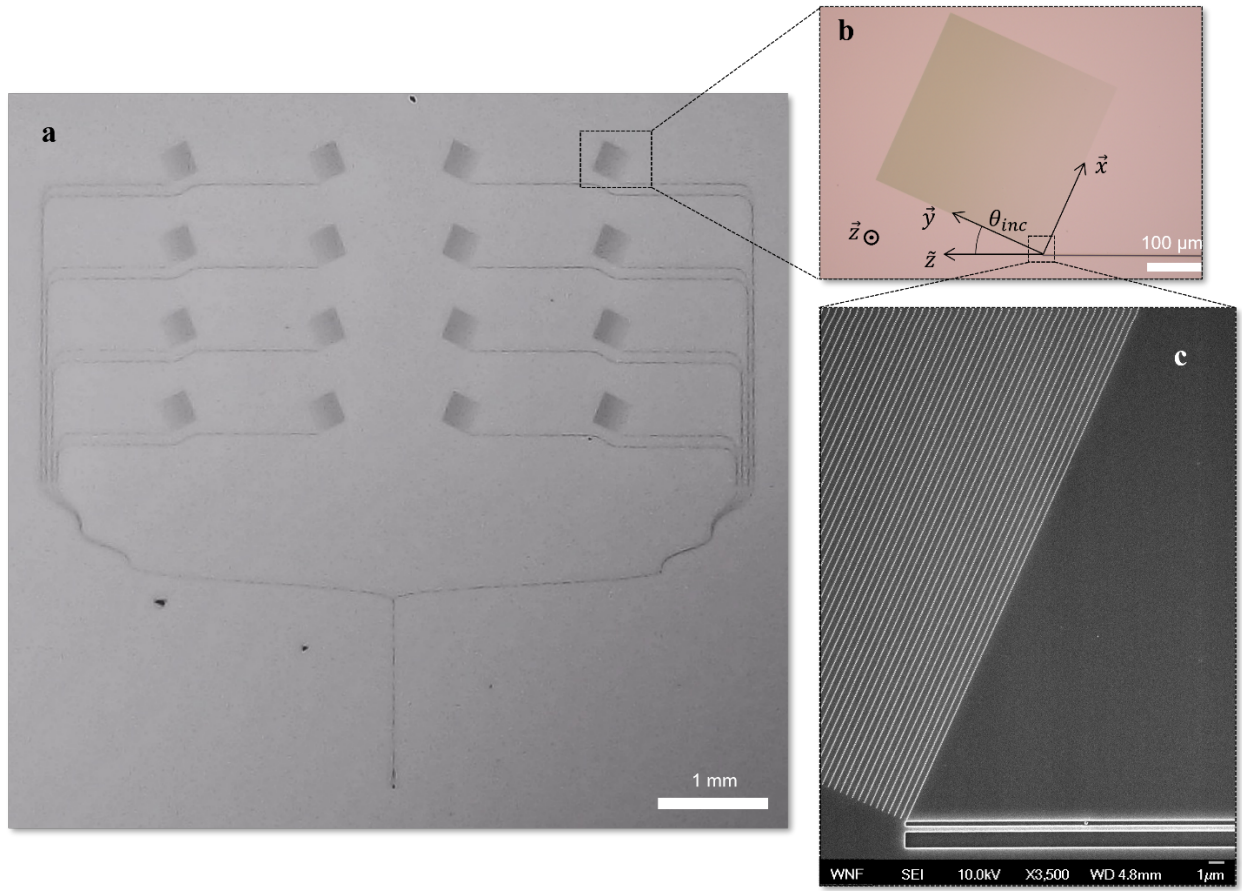

**Figure 2**: **a** Optical microscope image of the PIC: 16 identical apodized gratings are fed lights using a single grating coupler. **b** Closeup view of the detail of one apodized grating and system of coordinates. **c** SEM picture of one fabricated apodized grating.

We characterize the hybrid photonic platform using a fiber coupled microscopy setup (details are in Methods and Supplementary materials). First, we present the results on the vortex beam generators with orbital angular momenta. The upper panel of Figs. 3a, 3b, and 3c shows the optical microscope images of the vortex phase profiles with $l$ of 0, 3 and 15, respectively. The corresponding measured intensities for the same three devices are collected in the design target focal plane (1mm above the meta-optics) and plotted in the lower panel of Fig. 3. The hyperboloid metalens ($l = 0$), designed for a numerical aperture (NA) of 0.15, exhibits a diffraction-limited focal spot with a full-width at half maximum (FWHM) of approximately 3 μm. The diffraction efficiency (method detailed in Supplement) of the meta-optics was measured at approximately 69.8%. Details of the measurement of efficiency are provided in Fig. S2

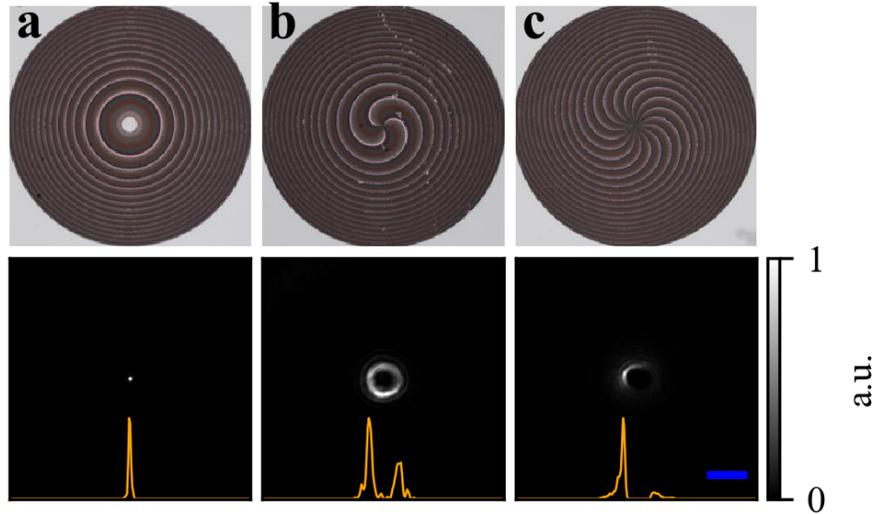

**Figure 3**: Optical microscope images (upper panel) and measured intensity in the focal plane of corresponding meta-optics (lower panel): **a** Hyperboloid converging lens, **b** vortex 3, **c** vortex 15. The common scale bar (in blue) measures 50 µm. The orange curves show the horizontal section cut of the 2D intensity.

We then compare the hyperboloid, the logarithmic-asphere, and the cubic phase meta-optics[21] in terms of their depth of focus. Figures 4a, b, and c show these three different meta-optics and their corresponding cross-sections in the left panel. As expected, both the logarithmic-asphere and cubic-phase mask exhibit a longer depth of focus compared to the hyperboloid metalens[22,23]. The hyperboloid lens has a depth of focus of 86 µm, while the logarithmic-asphere metalens exhibits a twofold increase in depth of focus. The logarithmic-asphere device also has a larger focal spot (FWHM ≈ 4.5 µm) than the hyperboloid (FWHM ≈ 3 µm). The depth of focus of the cubic meta-optics exceeds the boundaries of the measured depth range. The latter demonstrates a Gaussian central spot of approximately 5.8 µm FWHM and the direct neighboring peak intensity drops by 37% compared to the central one. The rays parallel to the chief ray are due to the cubic-phase inherent distortion. Essentially, the cubic phase generates an accelerating Airy beam[24,25].

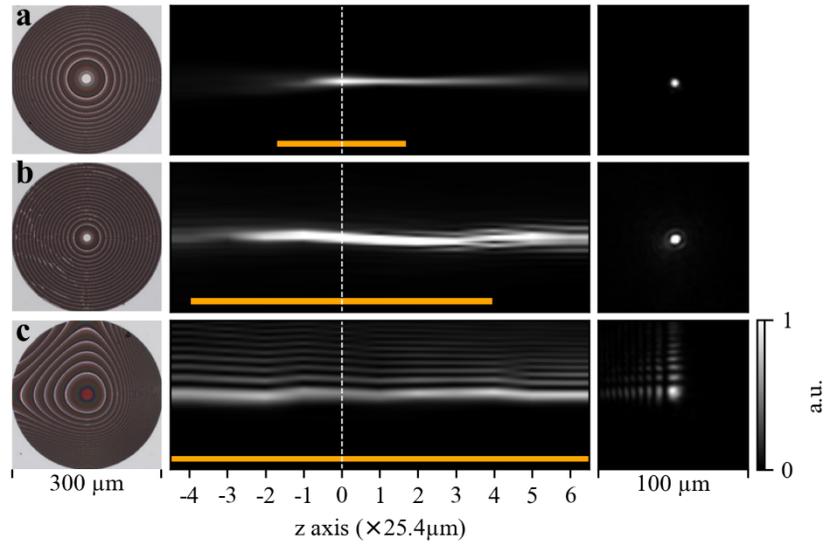

**Figure 4**: **Characterization of meta-optical lenses with different depths of field.** **a** hyperboloid metalens, **b** logarithmic-asphere metalens, **c** cubic-phase profile meta-optics. The left panel shows the corresponding optical microscope images. The middle panel shows the cross sections of the propagated beam intensity along the optical axis z. Each cross section is a concatenation of 11 planes along the optical axis, spaced at 25.4 µm, with linear interpolation. The 0 coordinate refers to the design focal plane location (at 1 mm above the PIC) and the corresponding intensities at this plane are shown in the right panel. The orange bar depicts the depth of focus for each device.

Finally, we characterize a set of three different holograms fabricated on the same chip. The upper panel of Fig. 5a, b, and c shows optical images of their phase masks. Binary intensities of the target holograms are shown in the middle panel and the intensity of the fabricated holograms are plotted in the lower panel. We observe a good fidelity between the holograms and their corresponding ground truth. According to the peak signal-to-noise ratio (PSNR), the "faceprint" hologram (Fig. 5b) shows the highest fidelity with a value of 11.4 while the "seal" and the "W" hologram rank second and third with PSNR values of 5 and 4.7, respectively. A possible explanation for the higher score of the "faceprint" hologram may be related to its smaller size. As in our Gerchberg-Saxton algorithm[26], the angular spectrum propagation method[27] is based on fast Fourier transforms of the fields, the size of the field impacts the higher spatial cut-off frequencies, causing the retrieved phase of the larger holograms to lose more bandwidth.

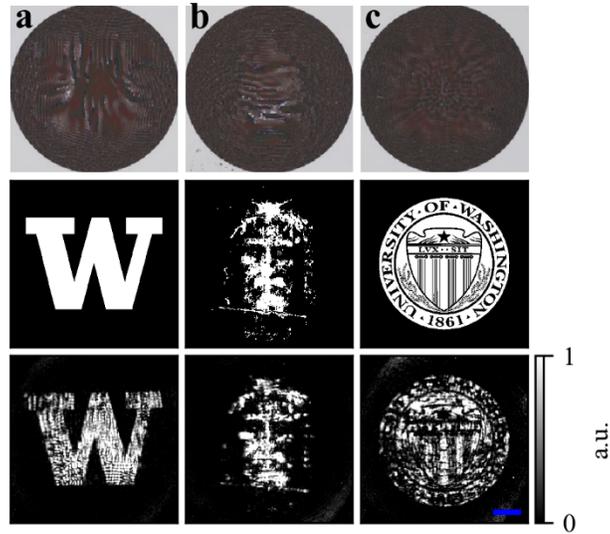

**Figure 5: Characterization of the holograms. a** Letter W. **b** Faceprint. **c** Seal of the University of Washington. Optical microscope images, target hologram and measured hologram in the designed projection plane are shown in upper, middle, and lower panels, respectively. The common scale bar (in blue) measures 50 μm.

# Discussion

We exploited meta-optics' ability to shape optical wavefronts to create a multifunctional interface between integrated photonics and free-space optics. While all the light beams coming out of the PIC are identical, by placing a different meta-optics on top of each grating, we can manipulate light differently. Specifically, the demonstration of holography provides an opportunity to engineer the beam shape to any desired functionality. If the light beam coming out of the PIC has some non-idealities due to fabrication imperfections, an additional meta-optics can be designed and fine-tuned to reshape the beam. Furthermore, meta-optics can provide an additional momentum vector to different beams, allowing light beams coming from gratings at different locations on the chip to form an ordered array of beams in free space. Such ability to reshape multiple beams can be particularly useful for active photonics. By placing the active components far from each other in a PIC, we can reduce the energy density and routing complexity of control circuits. However, by using a separate layer of passive meta-optics we can aggregate these beams in an ordered array. Interestingly, we found the alignment accuracy of the meta-optics and PICs can be on the orders of ~ 10 μm, making the co-packaging of these devices significantly simpler. We envision our demonstrated hybrid PIC/meta-optics platform will allow combining the best of both technologies: large space bandwidth product and active functionalities exploiting strong light-matter interaction, with far-reaching impact on the field of optical information science (both classical and quantum) and imaging.

## Materials and Methods

### Design

The design and fabrication of the PIC follows previous reports[20]. A standard straight-trench grating coupler was designed to couple an optical fiber to the PIC[5]. The grating coupler is connected to a network of waveguides splitting into 16 lines to irrigate the distal photonic devices. Each waveguide is terminated by a tapered profile to leak the waveguide mode into a one-dimensional collimated Gaussian slab mode. The waveguide-to-slab mode conversion step was accomplished using a fully etched waveguide, and the coupling distance between the waveguide and the slab was tapered from 1 µm to 0.29 µm over a length of 1000 µm to generate the tilted Gaussian slab mode incident on the apodized grating. Then, each apodized grating outcouples the slab mode into a free-space 2D Gaussian mode. The apodized gratings were partially etched over 60 nm deep and designed to generate a waist of 150 µm. The components of the photonic chip were designed for a wavelength of 780 nm using a 220 nm thick $Si_3N_4$ platform encapsulated with $SiO_2$-like refractive-index materials.

Both analytical and numerical methods were used for the design of the meta-optics. First a library of meta-atoms was generated to maximize the transmitted intensity and achieve a $2\pi$ phase coverage at the designed wavelength[28]. The details of the meta-atoms and the phase-to-pillar size trend are provided in Fig. S3. The meta-optics exhibiting focusing behavior were designed using analytical spatial phase profiles. The metalens, and the two vortex phase distributions are given by:

$$\varphi(r,\theta) = \frac{2\pi}{\lambda}\left(f - \sqrt{r^2 + f^2}\right) + l\theta$$

While $l = 0$ gives the phase profile for the hyperboloid lens, the vortex beams are designed with orbital angular momentum of $l = 3$ and 15, respectively. $r$ and $\theta$ represent the cylindrical coordinates of the phase mask and $f$, the focal length, is set at 1 mm. The spatial phase profile of the logarithmic-asphere meta-optics is given by[21]:

$$\varphi(r) = \int_0^r \frac{r dr}{\sqrt{r^2 + \left[s_1 + (s_2 - s_1)\left(\frac{r}{R}\right)^2\right]^2}}$$

where R is the aperture radius; $s_1$ and $s_2$ denote two ends of the depth of focus. We chose $s_1$=400 μm and $s_2$=1500 μm for our case. The cubic phase profile is given by:

$$\varphi(r) = \frac{2\pi}{\lambda}\left(f - \sqrt{r^2 + f^2}\right) + \frac{\alpha}{R^3}(x^3 + y^3)$$

where the parameter α denotes the strength of the cubic term and is chosen as 58π and $r^2 = x^2 + y^2$. In all these lenses R is kept at 150 μm.

The phase profiles of the three holograms were retrieved using a modified Gerchberg-Saxton algorithm along with an angular spectrum method to project the hologram at a given plane above the meta-optics slab. For the sake of simplifying the characterization, the projection planes of the all the meta-optics were design at a distance f=1mm on the z axis. The phase masks were then converted to a grid of $Si_3N_4$ meta-atoms arranged on a square lattice (as shown in Fig. S3).

**Fabrication process**

The PIC was fabricated in two steps. First, a 220 nm $Si_3N_4$ on buffer oxide chip was spin coated with a 400 nm thick layer of ZEP-520A electron-beam lithography (EBL) resist, and a 100 kV EBL system (JEOL JBX6300FS) was used to define the waveguide grating coupler, 330 nm wide waveguides, and slab mode coupling waveguides. The fully etched pattern was transferred using a dry reactive ion etch (RIE) consisting of $CHF_3/O_2$ chemistry. A second EBL step with a 150 nm thick layer of the same resist was then used to define the apodized gratings, which were then partially etched to a target depth of 60 nm. The entire chip was later spin coated with a thick layer of PMMA in order to conveniently mimic the refractive index of $SiO_2$.

Meta-optics were fabricated on a 500 μm thick 1" × 1" fused silica chip using a standard nanofabrication process. A layer of 780 nm of $Si_3N_4$ was deposited by PECVD at 350°C. A 300 nm thick layer of ZEP 520A followed by a thin film of anti-charging polymer (DisCharge H2O) were spin coated on the $Si_3N_4$. The resist was then exposed using the same EBL tool at 8 nA to form the nano-scatterer patterns. The resist was developed for 2 min in amyl acetate after removal of the conductive polymer in water. A 70 nm thick $Al_2O_3$ hard mask was evaporated and lifted-off in a bath of 1-methyl-2-pyrrolidone (NMP) at 90°C overnight followed by a short ultrasonication in dichloromethane (DCM). Due to the short distance between the apodized gratings and the optical fiber coupling on the PIC, and because the meta-optics must be aligned to the gratings, the fused silica chip was diced to make the patterns closer to the edge to leave sufficient space to approach the optical fiber. To form the nano-pillars, the patterns were transferred using a fluorine-based RIE process leaving a total thickness of 10 nm of $Al_2O_3$ on top of 778 nm of $Si_3N_4$. For the RIE step, the chips were bonding to a Si carrier wafer using a vacuum oil for thermal contact spread underneath the entire surface of the chip. Finally, the chip was unbonded in acetone and descummed in a gentle $O_2$ plasma. Additional SEM pictures of the holograms are given in Fig. S4.

## Characterization

To characterize the compound platform, a custom microscope composed of a long working distance objective lens aligned to a plano-convex lens with a visible camera is mounted onto an XYZ stage above the platform. A warm LED light delivered from an optical fiber, collimated, and split through a beam-splitter cube inserted between the objective and the plano-convex lens is used for alignment and imaging purposes. The meta-optics slab was mounted onto a 5 degree-of-freedom stage (XYZ + pitch + roll) for precise alignment to the photonic chip. The photonic chip was fixed to a 4 degree-of-freedom stage (XYZ + yaw) for fine adjustment purposes. Details of the optical setup are provided in Fig. S5.

## Acknowledgement

The research is supported by OIA-2134345. Part of this work was conducted at the Washington Nanofabrication Facility / Molecular Analysis Facility, a National Nanotechnology Coordinated Infrastructure (NNCI) site at the University of Washington with partial support from the National Science Foundation via awards NNCI-1542101 and NNCI-2025489.

# Supplementary information for

# Multi-functional interface between integrated photonics and free space

Quentin A. A. Tanguy[1, §], Arnab Manna[2, §], Saswata Mukherjee[1, §], David Sharp[2, §], Elyas Bayati[1], Karl F. Böhringer[1,3], and Arka Majumdar[1,2]

[1]Department of Electrical and Computer Engineering, University of Washington, Seattle, WA, USA

[2]Department of Physics, University of Washington, Seattle, WA, USA

[3]Institute for Nano-engineered Systems, University of Washington, Seattle, WA, USA

[§]Equal contribution

* Corresponding author: arka@uw.edu

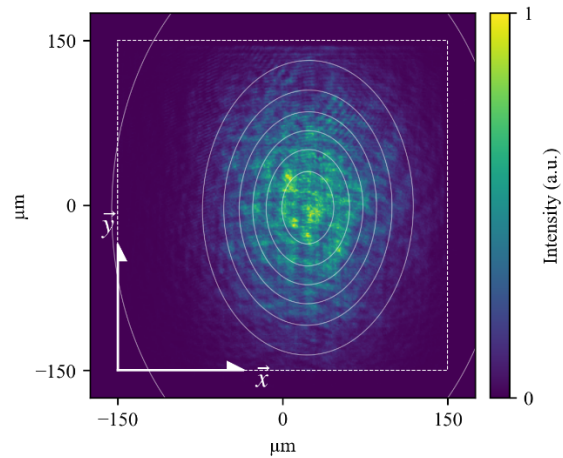

**Figure S1. Characterization of apodized grating and Gaussian fit.** To characterize the beam profile emitted by the apodized gratings, a 2D Gaussian profile was fitted on the intensity collected in the plane of the PIC using a custom microscope setup of magnification 2.9 with no meta-optics above it. The aperture of the apodized grating is depicted by the dashed white square in the figure. The concentric ellipsoids are the isoline of the contour of the 2D Gaussian fit.

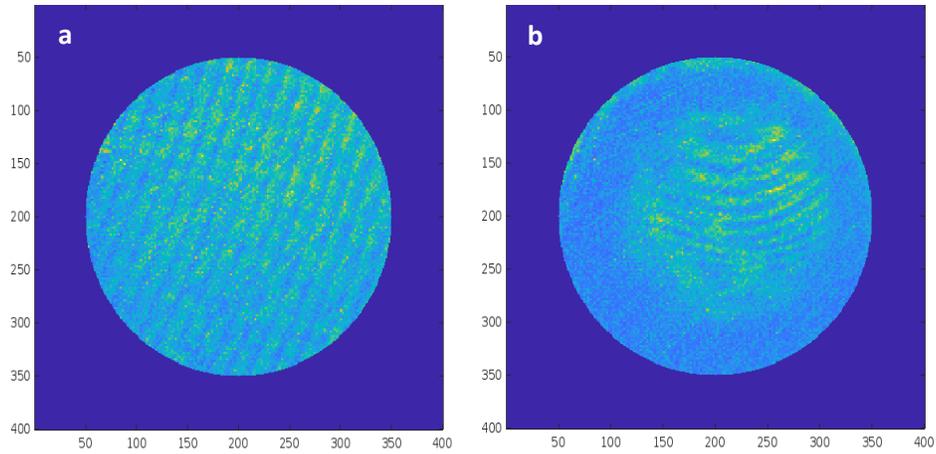

**Figure S2. Transmission efficiency calculation. a** Measurement of the light intensity passing through the glass slab without meta-optics. **b** Measurement of the hyperboloid metalens at an off-focus plane to calculate the total transmitted intensity. The sum of the individual pixels of the first configuration is divided by that of the second configuration to estimate the transmission efficiency of the metalens. The resulting efficiency gives approximately 69.8%.

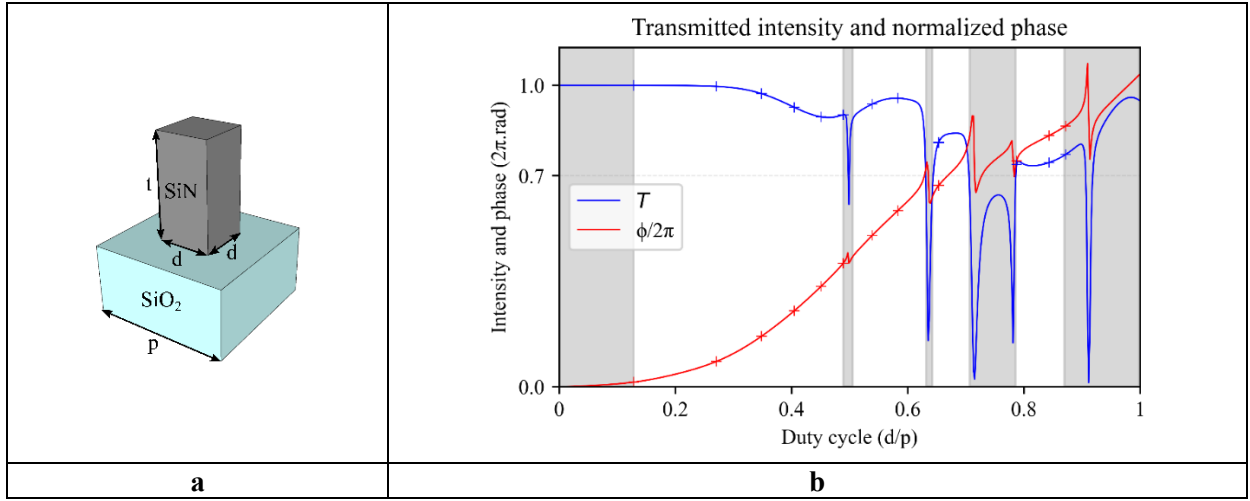

**Figure S3. Transmitted intensity and phase selection from RCWA simulations. a** The meta-atoms are made of $Si_3N_4$ square nanopillars of size $d$, height $\lambda$ and lattice $p = 0.7\lambda$ on fused silica[9]. **b** Numerical simulations using Rigorous Coupled-Wave Analysis (RCWA) were used to estimate phases and transmissions achievable by the selected nano scattering structure. Finally, 12 phase levels were selected between 70 nm and 476 nm, avoiding extreme duty cycle regions to prevent the risk of fallen pillars and merged patterns during fabrication. Areas grayed out correspond to high resonance regions and were avoided. Only intensities above 70% of transmission were kept. Discrete transmissions and phases corresponding to the selected pillar sizes are depicted by crosses.

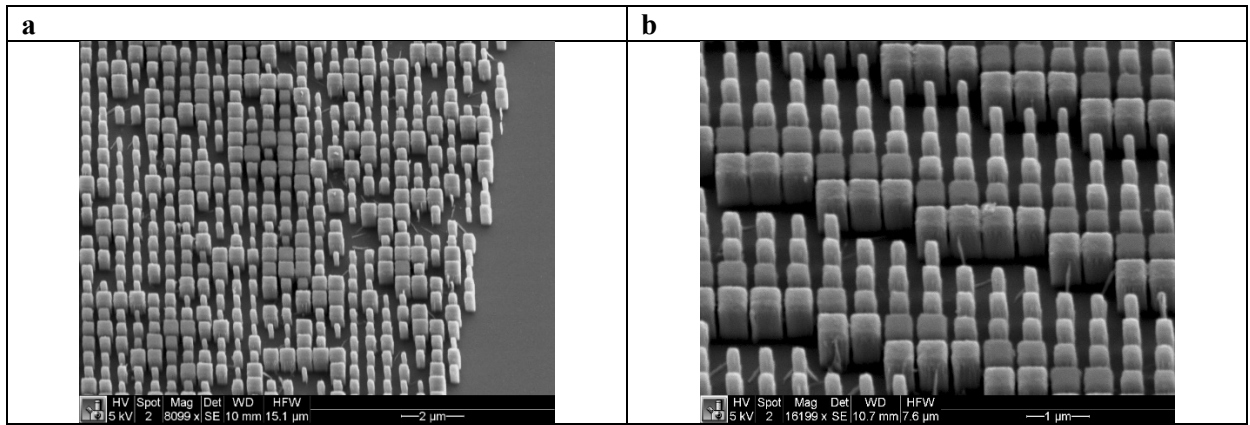

**Figure S4. SEM pictures of the meta-optics. a** Hologram. **b** Logarithmic-asphere.

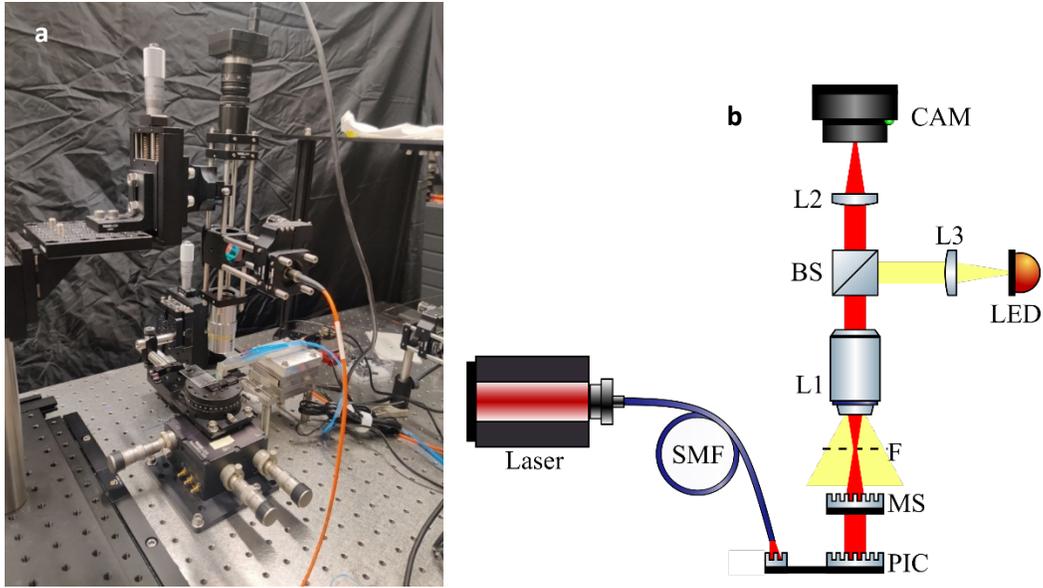

**Figure S5. Optical characterization setup. a** Picture of the setup. **b** Schematic. SMF: single-mode fiber, MS: metasurface chip, F: focal plane, L1: objective lens (Mitutoyo 10x plan Apo ∞), BS: beam splitter cube (Thorlabs CCM1 BS014), L2: plano-convex lens, L3: plano-convex lens. CAM: camera (Point Grey CMLN 13S2M CS). A warm LED is used for alignment and turned off for measurements.